\documentstyle[12pt,aasms4]{article}

\newcommand{\beq}{\begin{equation}}
\newcommand{\eeq}{\end{equation}}

\newcommand{\etal}{et al.}

\def\rso{R_\odot}

\def\llso{\log\,L/L_\odot}
\def\mso{M_{\odot}}
\def\msoy{\mso \, {\mathrm yr}^{-1}}
\def\kms{\hbox{\, km\,s$^{-1}$}}

\def\ra{\rightarrow}
\def\simgr{\,\hbox{\hbox{$ > $}\kern -0.8em \lower 1.0ex\hbox{$\sim$}}\,}
\def\simle{\,\hbox{\hbox{$ < $}\kern -0.8em \lower 1.0ex\hbox{$\sim$}}\,}

\def\ra{\rightarrow}
\def\bn{\bigskip\noindent}

\lefthead{Langer, Garc\'{\i}a-Segura, and Mac Low}
\righthead{Formation of Homunculus around $\eta$ Car}

\begin{document}

\title{
Giant Outbursts of Luminous Blue Variables and the Formation of 
the Homunculus Nebula Around $\eta$~Carinae
}

\author{Norbert Langer\altaffilmark{1,2}}
\affil{Max-Planck-Institut f\"ur Astrophysik, Postfach 1523,
D-85740 Garching, Germany; ntl@astro.physik.uni-potsdam.de}

\author{Guillermo Garc\'{\i}a-Segura}
\affil{Instituto de Astronomia - UNAM, Apdo. Postal 70-264,
04510 Mexico D.F., Mexico; ggs@astroscu.unam.mx}

\and 
\author{Mordecai-Mark Mac Low\altaffilmark{3,4}}
\affil{Max-Planck-Institut f\"ur Astronomie, K\"onigstuhl
17, D-69117, Heidelberg, Germany; mordecai@mpia-hd.mpg.de}

\altaffiltext{1}{present address:
Institut f\"ur Physik,
Universit\"at Potsdam, D-14415 Potsdam, Germany}
\altaffiltext{2}{Also: Instituto de Astronomia - UNAM, Apdo. Postal 70-264,
04510 Mexico D.F., Mexico }
\altaffiltext{3}{Also: Department of Astronomy, University of Illinois at
Urbana-Champaign, and Department of Astronomy and Astrophysics,
University of Chicago}
\altaffiltext{4}{present address: Department of Astrophysics, American
Museum of Natural History, 79th Street at Central Park West, New York, 
NY, 10024-5192, USA}

\begin{abstract}
The observed giant outbursts of Luminous Blue Variables (LBVs)
may occur when these massive stars approach
their Eddington limits.  When this happens, they must reach a
point where the centrifugal force and the radiative acceleration
cancel out gravity at the equator.  We call this the $\Omega$-limit.  When
stars are close to the $\Omega$-limit, strong non-spherical mass loss should
occur.  This suggests a scenario where a slow and very dense wind, 
strongly confined to the equatorial plane, is followed by a fast
and almost spherical wind.
We compute two-dimensional hydrodynamic models of the evolution of 
the nebula formed from such interacting winds, using
parameters consistent with the outburst of
$\eta$~Carina in the last century. 
This outburst
gave birth to the Homunculus, the hourglass-shaped inner part of a
highly structured circumstellar nebula.  
Assuming the star was very close to the $\Omega$-limit during
outburst, our models produce gas distributions 
that strongly resemble the Homunculus on large and small scale. 
This supports the general conjecture that giant outbursts in LBVs
occur when they approach the Eddington limit.
Our models constrains the
average mass loss rate since the outburst to values 
smaller than  the present-day mass loss rate and suggest that $\eta$~Car
is approaching another outburst.
Our models imply that the occurrence of giant LBV outbursts depends
on the initial stellar rotation rate, and that 
the initial angular momentum is as important to
the evolution of very massive stars as their initial mass
or metallicity.

\end{abstract}

\section{Introduction}
Current stellar evolution models predict that Galactic stars
initially more massive than 25--30$\mso$ lose more than 50\% of their
initial mass, and stars above 30--35$\mso$ more than 80\% 
(Maeder 1992, Woosley, Langer \& Weaver 1993). 
\markcite{1} \markcite{2}
Much of this mass loss is thought to 
occur during a short-lived and highly unstable stage preceding the
Wolf-Rayet stage, that may be observed in the form of the luminous
blue variables (LBVs; cf. Maeder 1989, Pasquali et al.\ 1997),
which are located close to the Humphreys-Davidson (HD) limit
in the HR~diagram, beyond which no normal stars are observed
(Humphreys \& Davidson 1979).
It is strongly debated what produces the giant LBV outbursts
observed in these stars (Langer et al.\ 1994, 
Nota \& Lamers 1997).\markcite{lan94}\markcite{nl97}
Among the potential mechanisms 
(see Humphreys \& Davidson 1994, and references therein),
the idea that massive stars reach their Eddington luminosity close to the
HD~limit is particularly appealing since the Eddington limit
appears to be located very close to the HD~limit in the HR diagram
(cf. Davidson 1971, 
Lamers \& Noordhoek 1992).

Time-varying massive star
winds are able to produce circumstellar nebulae with kinematic properties
similar to those observed around LBVs (Frank, Balick \&
Davidson\markcite{18}
1995; Nota et al.\markcite{10} 1995; Garc\'{\i}a-Segura, Mac~Low \& Langer
\markcite{19} 1996, \markcite{frd98}Frank, Ryu, \& Davidson 1998).
\markcite{how97}
The morphology of the nebula around the extraordinary star $\eta\,$Carinae
may give essential clues for the understanding of these giant outbursts.
During its eruption
from 1840 to 1860~A.D., $\eta\,$Car --- today a telescopic
object --- was the second brightest star in the sky\markcite{4} (van
Genderen \& The 1984).  Recent observations by the {\em Hubble Space
Telescope} have revealed in spectacular detail the resulting
circumstellar nebula \markcite{7}\markcite{m98}(Humphreys \& Davidson
1994, Morse et al. 1998), the
hourglass-shaped inner part of which is known as the
Homunculus\markcite{8} \markcite{9}(Meaburn, Wolstencroft \& Walsh
1987, Allen 1989).  This is now the best studied example of the
bipolar structures often observed around LBVs
such as $\eta\,$Car\markcite{10} 
(Nota \etal\ 1995).  


\section{Winds of rotating stars near the Eddington limit}
%
It has been proposed 
(e.g. Maeder 1989) \markcite{mae89}
that sufficiently luminous stars, after core hydrogen exhaustion,
may arrive at or exceed
their Eddington limit 
\begin{equation}
\Gamma \equiv L/ L_{\rm Edd} = 1,
\end{equation}
as their Eddington luminosity $L_{\rm Edd}=4\pi cGM/\kappa$
drops below their actual luminosity~$L$ as the
opacity coefficient~$\kappa$ increases.
The theory of radiation-driven stellar winds predicts that the mass loss
rate will increase as $\dot M \propto (1-\Gamma )^{-\mu}$
with $\mu> 0$ for stars approaching the Eddington limit \markcite{cas75}
(cf. Castor, Abbott \& Klein 1975).  This mass loss rate formally diverges
for $\Gamma \ra 1$ (though see Owocki \& Gayley 1997), suggesting that the
strong mass loss \markcite{owo97}
associated with $\Gamma\simeq 1$ may be related to giant outbursts of LBVs.

However, rotation reduces the luminosity required for 
all external forces to balance each other at the
stellar surface (Langer 1997). \markcite{lan97}
The Eddington limit 
$\Gamma < 1 $ should then be replaced by a criterion that we will
call the $\Omega$-limit, 
\begin{equation}
\Omega \equiv v_{\rm rot}/v_{\rm crit} < 1,  
\end{equation}
with $v^2_{\rm crit} \equiv  v^2_{\rm
esc} /2 = GM(1-\Gamma )/R$, $M$ 
and $R$ being the stellar mass and radius, and $v_{\rm esc}$ being the
polar escape velocity. 
$\Omega=1$ implies that centrifugal and radiation force balance gravity
at the equator, while at higher latitudes gravity still dominates.
The feedback of rotation on the local surface luminosity is neglected here, 
since according to the generalized von Zeipel theorem (Kippenhahn 1977)
the radiation flux at the equator may be either reduced or enhanced,
depending on the internal rotation law; this may have been overlooked 
in the recent criticism of the $\Omega$-limit by Glatzel (1998).
Here, we apply the result of Friend \& Abbott (1986) that the
mass-loss rate of rotating hot stars depends on $\Omega$ as
$\dot M \propto (1-\Omega )^{-\nu}$ with \markcite{26}
$\nu\simeq 0.43$.  

We divide the evolution of 
a star approaching the $\Omega$-limit --- that is, going through an
outburst --- into three
phases. In the first phase, before the star reaches the $\Omega$-limit
($\Omega < 1$), 
it has the
fast, energetic wind expected of a luminous blue star. In the second phase,
reaching the $\Omega$-limit ($\Omega \simeq 1$) 
has three consequences
for the wind: the mass loss rate is much higher than before;
the mass flux increases strongly at latitudes close
to the equator; and
the bulk of the wind is slow since the equatorial escape
velocity is almost zero.  In the third phase the outburst is over, the
stellar radius has decreased, and the configuration is similar to the
first phase. The smaller value of $\Gamma$ before and after the
outburst has two consequences: larger wind velocities due to larger
escape velocities {\em and} smaller values of~$\Omega$ leading to more
spherical winds.

To compute the latitudinal dependence of the wind properties of a star
close to critical rotation ideally requires multi-dimensional models of
the star and its outflowing atmosphere, which are not available.
However, Langer (1997, 1998) argued that the stellar flux and the radius
might still vary only weakly from pole to the equator in very luminous stars.
Therefore, we applied equations similar to those found by Bjorkman
\& Cassinelli (1993, BC) for winds of rotating stars in the limit
of large distance from the star:
\beq
v_{\infty}(\theta) =  \zeta v_{\rm esc} \left(1 - \Omega\,
     \sin\theta  \right)^{\gamma} \,\, ,
\eeq
\beq
(4 \pi r^2\rho)_{\infty} (\theta) =
     {\alpha\over 2} \delta \dot M_0 \left(1 - \Omega\,
     \sin\theta      \right)^{\xi} / v_{\infty}(\theta)
     \,\,   ,
\eeq
where we set the parameters defined in BC to $\zeta=1$ ,$\gamma=0.35$,
and $\xi=-0.43$.
The correction factor $\delta$ is introduced to ensure that the total
stellar mass loss rate $\dot M_0$ obeys
$\dot M_0 =  \int v_{\infty}(\theta) \rho (\theta) \sin\theta 
\,\,d\theta\, d\phi$
at the inner boundary of our grid (cf. Table~1).
$v_{\infty}$ is the terminal wind velocity, and
$(4 \pi r^2\rho)_{\infty}$ the terminal wind density times
$4 \pi r^2$, as function of the polar angle $\theta$.
The quantity $\alpha$ is defined by
\beq
\alpha = \left( \cos\phi' + {\cot}^2 \theta
    \left( 1 + \gamma { \Omega\, \sin\theta \over
     1 - \Omega\, \sin\theta } \right) \phi'\, \sin\phi'
    \right)^{-1}   \,\,,
\eeq
with
$\phi' = \Omega\, \sin\theta v_{\rm crit} /
             ( 2\sqrt{2} v_{\infty}(\theta ) )$.
Eq. (5) differs from the corresponding quantity defined by BC 
in their implicit formula (26) (cf. also Owocki et al. 1994, 
Ignace et al. 1996). This difference came along
originally through a misinterpretation of BC's equations,
i.e., in equations (3) to (5) and in the equation defining 
$\phi'$, the $\theta$s were taken as $\theta_0$s, the initial
co-latitude of the streamline. 
We note that since for wind compressed zone models near the equator
it is $\theta_0 \simeq \theta$, our models are similar 
to wind compressed zone models when $\Omega \simeq \Omega_{\rm th} \simeq 1$,
where $\Omega_{\rm th}$ is the threshold value for the formation of a
wind-compressed disk.
Since $\phi' < \pi/2$ for $\Omega \le 0.995$, our formulation has the
advantage of avoiding the formation of wind compressed disks for large
$\Omega$, a structure which can not be numerically resolved in our 
calculations, whose properties cannot be well predicted, and whose very
existence has even been questioned (Owocki \etal 1994, 1996). At the same time,
wind density and velocity distributions obtained from our approach are
similar to those derived from the formalism of BC,
provided that $\Omega_{\rm th} \approx 1$ and we choose 
$\Omega_{\rm BC} = \Omega \Omega_{\rm th}$, 
where $\Omega_{\rm BC}$ is the value to be inserted in BC's equations.
We shall see that the exact nature of
the latitude dependence of the wind properties is not essential
for our main results, as long as a dense wind with enhanced mass loss
rate close to the equator occurs between two phases of an energetic,
more or less spherical wind.

We simulate the LBV outburst phenomenon by 
assuming the wind properties to be constant during each phase. 
For the pre-outburst wind, which is only used to initialize the
numerical grid and to which our results are insensitive, we took a
mass-loss rate of $\dot M=10^{-3}\msoy$, a wind final velocity
$v_{\infty}=450\kms$ and $\Omega =0.53$.  For the post-outburst wind,
we used two different sets of parameters.  The one that we prefer
appears to reasonably reproduce the observed morphology of the
Homunculus, with $\dot M=1.7\times 10^{-4} \msoy$,
$v_{\infty}=1800\kms$, and $\Omega = 0.13$.  For comparison we also
computed a model that represents $\eta$~Car's {\it presently} observed
wind parameters\markcite{31} of $\dot M = 3\times 10^{-3}\msoy$,
$v_{\infty} = 800\,$ km~s$^{-1}$, and $\Omega = 0.3 $
(Davidson \etal\ 1995).  The high wind velocity of our preferred
model, $v_{\infty}=1800\kms$, corresponds to a stellar radius $R\simeq
21\rso$ or $\log T_{\rm eff} \simeq 4.7$ for an O~star wind with
$\zeta =3$ at $\llso = 6.4$, implying that the star
strongly contracted after the episode of mass-loss.

The parameters of the outburst wind largely determine the morphology
of the resulting nebula.  To compute these parameters, we assumed
the following stellar properties:
$M=80\mso$, $\llso =6.4$, and $\log T_{\rm eff}=4.2$, implying
$R=210\rso$, in agreement with observational estimates (cf. Humphreys
\& Davidson 1994).  The final wind velocity 
follows from the escape velocity for a specified value of $\Gamma$.
For our preferred model we took a mass loss rate of
$\dot M =7\times  10^{-3}\msoy$ and obtained a Homunculus mass of
roughly $0.15\mso$ (van~Genderen \& The 1984), while for our model
assuming present-day wind parameters, we used an increased outburst
mass loss rate of $5\times  10^{-2}\msoy$ such that we obtained a 
nebula mass of $1\, \mso$ \markcite{7}(Humphreys \& Davidson 1994).
 
\section{Hydrodynamic models}

We perform two-dimensional 
hydrodynamic simulations of the wind interaction; first results have
already been reported by Garc\'{\i}a-Segura, Langer \& Mac~Low (1997).
We use the hydrocode ZEUS-3D\markcite{gar97}
developed by M. L. Norman and the Laboratory for Computational
Astrophysics.  ZEUS-3D is a finite-difference, fully explicit, Eulerian
code descended from the code described by \markcite{33}Stone \&
Norman (1992).  We used spherical coordinates for our simulations,
with a symmetry axis at the pole, and reflecting boundary conditions
at the equator and the polar axis.  See Garc\'{\i}a-Segura \etal\
\markcite{16} (1996) for further details about our numerical method.
Our models have grids of $200 \times 360$ zones, with a radial extent
of 0.125 pc, and an angular extent of $90^{\circ}$.  The innermost
radial zone lies at $r=9.7 \times  10^{15}\,$cm.  

We compute the hydrodynamic evolution of the circumstellar gas,
starting our computations at a time $t=1840\,$yr, and run them until
$t=1995\,$yr.  
Our outburst scenario leads to a characteristic distribution of the
circumstellar gas.  The initial fast wind blows a stellar wind bubble
which forms the background for the subsequent development of the
nebula.  During outburst, the wind becomes slow and dense, and the
stellar rotation concentrates it toward the equatorial
plane \markcite{28}(BC; \markcite{ign96}
Ignace \etal\ 1996).  When the final fast wind starts in the center
of this nebula, it sweeps up the dense wind from the outburst into a
thin, radiatively cooled shell that fragments due to dynamical
instabilities\markcite{19} (Garc\'{\i}a-Segura \etal\ 1996).  The
shell expands more easily into the lower density wind at the poles,
producing a double-lobed structure, as shown in Figures~1 and~2.
 
In Figure~1, we show six models computed with various values of $\Omega$
and $\Gamma$ for the outburst wind given in Table~1, and otherwise
using the parameters of our preferred model.  We find three major
results: First, the nebular shape appears nearly independent of the
Eddington factor, $\Gamma$.  Second, the nebula is strongly confined
in the equatorial plane only for the case with nearly critical
rotation ($\Omega=0.98$).  Finally, in this case, we
obtain a structure very similar to that of the Homunculus, with two
almost spherical lobes and an equatorial density enhancement that
is an expanded relic of the outburst wind. 
%
%
In Figure~2 we show the results of a computation identical to that of
Figure~1a --- in particular with $\Omega =0.98$ --- but using twice the
radial resolution.  
We find it striking how well this
model, without much fine tuning, not only reproduces the large-scale,
bipolar morphology, but also the small-scale turbulent structure seen
in high-resolution observations of the Homunculus
\markcite{7}\markcite{m98}(Humphreys \& Davidson 1994, Morse et al.\ 1998).

We also compute a model at the same resolution as Figure~2 using
$\eta$~Car's {\it presently} observed wind parameters.
We find that the large scale shape of the resulting nebula, shown in 
Figure~3,
is almost identical to that shown in Figure~2, but the higher
wind densities cause the wind termination shock to be strongly
radiative. This changes the Vishniac
instabilities\markcite{32} (Vishniac 1983) seen in Figure~2 into
ram-ram-pressure instabilities\markcite{19} \markcite{34} (Vishniac
1994, Garc\'{\i}a-Segura \etal\ 1996), which have a much spikier
morphology and a shorter wavelength, inconsistent with the observed
structures.  
We emphasize that only the properties of the post-eruption
wind are responsible for this feature, not the larger shell mass
obtained in this case.
This result might imply that the current wind is not
representative of the wind over the last 140 years. Instead,
$\eta$~Car might have been smaller and hotter in the
recent past, with a post-outburst wind that has become slower and more
dense with 
time.  This is consistent with its gradual visual brightening over the
last 140 years\markcite{7} (Humphreys \& Davidson 1994), and suggests
that it is evolving towards another giant eruption.

\section{Discussion}
Our work extends previous hydrodynamic models for bipolar LBV
nebu\-lae\markcite{10}\markcite{18} (Nota \etal\ 1995, Frank \etal\
1995, 1998, Dwarkadas \& Balick 1998) by relating the outburst to the
properties of evolving, massive, post-main sequence stars. 
In contrast to Nota \etal\ (1995) and Frank \etal\ (1995), 
who concluded that a strong equatorial density enhancement must
have existed before the outburst occurred, we obtain
the two lobes, including their small-scale structure,
and the equatorial density enhancement
self-consistently as a consequence of the evolutionary state of the star.
Frank \etal\ (1998)\markcite{f98} used an arbitrary non-spherical wind
during the 
post-outburst phase to produce a bipolar nebula.  However, such a wind
will only produce a bipolar shape if the wind termination shock is
strongly radiative and therefore momentum conserving, a condition we
have shown to be inconsistent with the small-scale morphology of the
Homunculus.  \markcite{db98}Dwarkadas \& Balick (1998) introduce instead a
ring-like density distribution, again without relating it to the
underlying star. 

Our result appears to be quite general, because all stars, even
slow rotators,  must by definition
arrive at critical rotation if they approach their Eddington limit.
The strong dependence of the nebula shape on $\Omega$ shown in Figure~1,
as well as the clear bipolar nature of virtually all LBV
nebulae\markcite{10} (Nota \etal\ 1995), lends strong support to the
idea that LBV's are stars approaching their Eddington limits that
reach critical rotation and lose large amounts of mass quickly.
%
In fact, a general mechanism for giant LBV outbursts is needed 
in order to understand the absence of stars beyond the
Humphreys-Davidson limit in the HR diagram, and the bipolar nature
of most LBV nebulae. Therefore, even though bipolar nebulae may also
form from interacting binary stars (e.g. Han, Podsiadlowski \& Eggleton 
1995), \markcite{han95}
and binarity has been
repeatedly proposed also for $\eta$~Car (van Genderen, de Groot \& The 1994,
Damineli, Conti \& Lopes 1997),\markcite{van94}\markcite{dam97}
it appears useful to continue to pursue single star models.
The conjecture that the Homunculus nebula around $\eta$~Car is a
paradigm rather than a freak has recently been supported by its strong
similarity to the nebula around the LBV HR~Carinae, as found
by Weis et al.\ (1997) and Nota et al.\ (1997). \markcite{wei97}
\markcite{not97} 
 
\acknowledgments NL is grateful to Gloria Koenigsberger for enabling
an extended visit to the UNAM Astronomy Institute, and to many
colleagues there for their extremely warm hospitality.  We thank
J. Fliegner, S. Owocki, M. Peimbert, S. White and in particular
J.~Bjorkman
for useful comments and discussions, and M. L.~Norman
and the Laboratory for Computational Astrophysics for the use of
ZEUS-3D. The computations were performed at the Pittsburgh
Supercomputing Center, the Supercomputer Center of the Universidad
Nacional Aut\'onoma de M\'exico and the Rechenzentrum Garching of the
Max-Planck-Gesellschaft.  This work was partially supported by the
Deutsche Forschungsgemeinschaft through grants La~587/15-1 and 16-1
and by the US National Aeronautics and
Space Administration.

\clearpage

\begin{center} {\Large Figure~Captions} \end{center}
%
\figcaption{Logarithm of the circumstellar gas density (in
g~cm$^{-3}$) in our hydrodynamic models at $t=1995\,$yr.  Figs.~1a-f
correspond to models a-f in Table~1.}

\figcaption{Same as Figure~1a, but using twice the radial resolution.
The resemblance of the Vishniac instabilities in the swept-up lobes
with the corresponding features in the high-resolution observations by
the Hubble Space Telescope (Humphreys \& Davidson 1994, Morse \etal\
1998) is striking.  The mass of the Homunculus nebula in this model
is~$\sim 0.15 M_{\odot}$ (van Genderen \& The 1984).}

\figcaption{Same as Figure~1a, but using $\eta$~Car's presently
observed wind parameters.  The higher mass-loss rate changes the
nature of the shell instabilities, as discussed in the text, producing
a morphology that disagrees with the observations.  The mass of the
Homunculus nebula in this model is $\sim 1
\mso$~(Humphreys \& Davidson 1994).}
\newpage
\begin{table}
\centering
\caption[junk]{Parameters during the outburst phase             
for the models shown in Figure~1.   $v_{\rm esc}$, the escape
velocity from the stellar surface in polar direction, is also the final
wind velocity in polar direction.  See text and figure captions for     
definitions of other quantities. 
}
\bn
\begin{tabular}{l|ccccc}
\hline
\hline 
 Model &$\Omega$&$\Gamma$&$v_{\rm rot}$&
     $v_{\rm crit}$&$v_{\rm esc}$   \\
  $\,$ &  $\,$  &  $\,$  &   $\kms $   &    $\kms $   &    $\kms $    \\ 
\hline 
 a    &  0.98  &  0.50  &  186.7      &   190.5      &    269     \\
 b    &  0.90  &  0.50  &  171.5      &   190.5      &    269   \\
 c    &  0.98  &  0.75  &  132.0      &   134.7      &    190   \\
 d    &  0.90  &  0.75  &  121.2      &   134.7      &    190   \\
 e    &  0.98  &  0.98  &   37.3      &    38.1      &     54   \\
 f    &  0.90  &  0.98  &   34.3      &    38.1      &     54   \\
\hline
\hline
\end{tabular}
\end{table}


\newpage
\begin{references}

\reference{9} Allen D. A., 1989, MNRAS, 241, 195 
\reference{27} Bjorkman J. E., \& Cassinelli J. P. 1993, 
     ApJ, 409, 429 ({\bf BC}) 
\reference{cas75} Castor J.I., Abbott D.C., Klein R., 1975, ApJ, 195, 157
\reference{dam97} Damineli A., Conti P.S. \& Lopes D.F., 1997, New
Astron., 2, 107
\reference{11} Davidson K., 1971, \mnras, 154, 415 
\reference{31} Davidson K., Ebbets D., Weigelt G., Humphreys R.M.,
Hajian A. R., Walborn N., \& Rose M., 1995, AJ, 109, 1784 
\reference{db98} Dwarkadas, V., \& Balick, B. 1998, AJ, 116, 829
\reference{18} Frank A., Balick B., \& Davidson K., 1995, ApJ, 441,
L77 
\reference{f98} Frank, A., Ryu, D., \& Davidson, K., 1998, ApJ, 500, 291
\reference{26} Friend D. B., \& Abbott D. C., 1986, ApJ 311, 701 
\reference{19} Garc\'{\i}a-Segura G., Mac Low M.-M., \& Langer N. 1996,
       A\&A, 305, 229  
\reference{gar97} Garc\'{\i}a-Segura G., Langer N.,  Mac Low M.-M., 
   1997, in {\em Luminous Blue
   Variables: Massive Stars in Transition}, ASP Conf. Ser. Vol.~120, 
    San Francisco,
    A. Nota, H.J.G.L.M. Lamers, eds., p.~332
\reference{} Glatzel W., 1998, A\&A, 339, L5
\reference{han95} Han Z., Podsiadlowski P., Eggleton P.P., 1995,
MNRAS, 272, 800 
\reference{20} Humphreys R. M., \& Davidson K. 1979, ApJ, 232, 409 
\reference{7} Humphreys R. M., \& Davidson K. 1994, PASP 106, 1025 
\reference{ign96} Ignace R., Cassinelli J.P., Bjorkman J.E., 1996, ApJ,
     459, 671
\reference{} Kippenhahn R., 1977, A\&A, 58, 267
\reference{13} Lamers H. J. G. L. M., \& Noordhoek R., 1993, in {\it 
Massive
       Stars: Their Lives in the Interstellar Medium}, J.~Cassinelli \&
       E.~Churchwell, eds., (San Francisco, ASP), ASP Conf. Ser.
       Vol.~35, p.~517 
\reference{lan97} Langer N., 1997, in {\em Luminous Blue
   Variables: Massive Stars in Transition}, ASP Conf. Ser. Vol.~120, 
    San Fransisco,
    A. Nota, H.J.G.L.M. Lamers, eds., p.~83
\reference{lan98} Langer N., 1998, A\&A 329, 551
\reference{16} Langer N., Hamann W.-R., Lennon M., Najarro F., 
Pauldrach
       A. W. A., \& Puls J. 1994, A\&A, 290, 819 
\reference{mae89} Maeder A. 1989, in {\em Physics of Luminous Blue
    Variables}, K.~Davidson et al., eds., Kluwer, p.~15
\reference{1} Maeder A., 1992, A\&A, 264, 105
\reference{8} Meaburn J., Wolstencroft R.D., \& Walsh J.R., 1987, A\&A,
181, 333 
\reference{m98} Morse, J., Davidson, K.,  Bally, J., Ebbets, M., 
Balick, B., \& Frank, A., 1998, AJ, 116, 2443
\reference{nl97} Nota A. \& Lamers H.J.G.L.M., (eds.), 1997, {\em Luminous 
Blue Variables: Massive Stars in Transition}, ASP Conf. Ser. Vol.~120, 
San Fransisco
\reference{10} Nota A., Livio M., Clampin M., \& Schulte-Ladbeck R., 1995, 
ApJ,
       448, 788 
\reference{not97} Nota A., Smith L., Pasquali A., Clampin M. \& Stroud M.,
      1997, ApJ 486, 338 
\reference{owo97} Owocki S. P., Gayley K.P.,
   1997, in {\em Luminous Blue
   Variables: Massive Stars in Transition}, ASP Conf. Ser. Vol.~120, 
   San Fransisco,
    A. Nota, H.J.G.L.M. Lamers, eds., p.~121
\reference{29} Owocki S. P., Cranmer S. R., \& Blondin J. M., 1994, ApJ,
424, 887 
\reference{owo96} Owocki S. P., Cranmer S. R., Gayley K.G., 1996, ApJ, 472, 
L115
\reference{pas97} Pasquali A., Langer N., Schmutz W., Leitherer C., Nota 
A.,
     Hubeny I., Moffat A.F.J., Drissen L. \& Robert C. 1997, ApJ, 478, 340
\reference{33} Stone J. M., \& Norman M. L. 1992, ApJS, 80, 753 
\reference{4} van Genderen A. M., \& The P.S ., 1984, Space
   Sci. Rev., 39, 317 
\reference{van94} van Genderen A. M., de Groot M.J.H., \& The P.S, 
    1994, A\& A, 283, 89
\reference{32} Vishniac E. T., 1983, ApJ, 274, 152 
\reference{34} Vishniac E. T., 1994, \apj, 428, 186
\reference{wei97} Weis K., Duschl W.J., Bomans D.J., Chu Y.-H. \& Joner 
M.D.,
   1997, A\& A, 320, 568
\reference{2} Woosley S. E., Langer N., \& Weaver T. A. 1993,  
  ApJ 411, 823
 
\end{references}
\end{document}